\documentstyle[eqsecnum,aps,prb,twocolumn,epsf]{revtex}
\newcommand{\PostScript}[3]{
\vspace{#1cm}
\begin{center}
\epsfysize=#2cm
\leavevmode
\epsfbox{#3}
\par
\end{center}
}

\begin{document}
\draft
\preprint{TPR-98-12}

\title{Ionic structure and photoabsorption in medium sized sodium clusters}

\author{S. K\"ummel$^1$, M.Brack$^1$, and P.-G. Reinhard$^2$}
\address{$^1$Institute for Theoretical Physics, University of Regensburg,
D-93040 Regensburg, Germany\\
$^2$Institute for Theoretical Physics, University of Erlangen,
D-91077 Erlangen, Germany}

\date{\today}
\maketitle
\begin{abstract}
We present ground-state configurations and photoabsorption spectra of
Na-7+, Na-27+ and Na-41+. Both the ionic structure
and the photoabsorption spectra of medium-size sodium clusters beyond
Na-20 have been calculated self-consistently with a nonspherical
treatment of the valence electrons in density functional theory. We
use a local pseudopotential that has been adjusted to
experimental bulk properties and the atomic 3s level of sodium.
Our studies have shown that both the ionic structure of the ground
state and the positions of the plasmon resonances
depend sensitively on the pseudopotential used in the calculation,
which stresses the importance of its consistent use in
both steps.

\end{abstract}


\narrowtext

Important progress has recently been made in the measurement of
photoabsorption spectra in small and medium-size sodium clusters
\cite{na41p,expna7p,mei}. This calls for a critical re-evaluation of the
theoretical description of the optical response of these systems.
The earliest theoretical investigations of alkali clusters
employed the jellium model \cite{ekardt1,beck} and were followed
by other groups \cite{brack1,yan1,montag1,guet1} (for a review,
see \cite{revmod}). The jellium model can
qualitatively explain many of the experimentally observed
features \cite{revmod}.
But a more realistic description of metal clusters is highly desirable, e.g.
since the unrealistic sharp
jellium edge gives rise to fundamental questions in the context of calculating
optical properties of metal clusters \cite{guet1,revmod}.

The all-electron {\it ab initio} methods of quantum chemistry treat the
 ionic degrees of freedom on the most sophisticated level
presently possible \cite{bonacic3,bonacic4}. But their computation times
grow so rapidly with the size that only small systems could be
studied so far.
Treating only the valence electrons explicitly and describing nucleus
plus core electrons by a pseudopotential leads to a considerable
simplification of the electronic system. The unrestricted three-dimensional
search for cluster ground-state configurations  using
Car-Parrinello \cite{carpar,rothlis} or Monte-Carlo methods with
non-local pseudopotentials \cite{martins1} is, however, still a task
of considerable complexity.

Several approximate methods have been developed for the study of the
ionic structure in larger clusters.
The ``spherically averaged pseudopotential scheme'' (SAPS)
\cite{iniguez,borstel}
optimizes the ionic positions in three dimensions while restricting
the density of the valence electrons to spherical symmetry; the
ion-valence electron
interaction is described by the simplest pseudopotential possible, the
Ashcroft empty-core potential. However, where
more accurate methods can be applied, these predict ground-state geometries
which differ considerably from the SAPS results.
The pseudopotential perturbation theory  and similar approaches
greatly improve
on the SAPS deficiencies in the treatment of the valence electrons, but
they either consider only the volume-averaged effects of ionic structures
\cite{serra,alasia}, or require their geometries as an input
\cite{ppstoer,bertsch}.
Approaching the problem from the opposite direction, a sophisticated
extension of the H\"uckel model \cite{spiegelmann} focuses on the
prediction of ground-state configurations without explicitly taking
the valence electrons into account.
Its results are in good agreement with {\it ab initio} calculations.
However, the parameters of this model must be adjusted to {\it ab initio}
calculations and the optical response of the electrons cannot be calculated.

With the ``cylindrically averaged pseudopotential scheme'' (CAPS) \cite{caps1},
a method has been developed that
allows one to self-consistently calculate the ground states of clusters
with several tens of atoms including ionic structure without restricting the
valence electrons to spherical symmetry. The ionic configuration is
hereby optimized by the method of simulated annealing that seems to be the
best method for coping with the strong isomerism found in
larger clusters. The electronic system is described in
density functional theory \cite{dreizler}; here we use the local density
approximation (LDA) with the functional of Perdew and Wang \cite{pw}.
By an interlaced iteration, the set of equations
\begin{displaymath}
\begin{array}{cc}
\frac{\partial E}{\partial \bf{R}}=0 \hspace{1cm}&
\frac{\delta E}{\delta n}=0
\end{array}
\end{displaymath}
is simultaneously solved self-consistently. Here $\bf R$
denotes the set of all ionic
positions, $n$ the electronic density and $E$ the energy
functional
\begin{displaymath}
\begin{array}{ccc}
E[n; {\bf R}]&=&T_s[n]+E_{\mathrm xc}[n]
+\frac{e^2}{2}\int \int \frac{n{{\bf(r)}n({\bf r'})}}{\left|{\bf
  r-r'}\right|} \,d^3 r' \, d^3 r
\nonumber \\  &&
+\int n({\bf r})V_{\mathrm ei}({\bf r; R}) \,d^3 r
+\frac{ Z^2 e^2}{2}\sum_{\stackrel{i,j=1}{i\ne j}}^N\frac{1}{\left| 
 {\bf R}_i-{\bf R}_j
  \right|}
\end{array}
\end{displaymath}
for a cluster of $N$ ions with $Z$ valence electrons each.

The efficiency of the scheme results from two approximations that are made
in the evaluation of the above energy functional. First, the interaction
between valence electrons and ionic cores is described by a local
pseudopotential:
\begin{equation}
V_{\mathrm ei}({\mathbf r; R}) =\sum_{i=1}^N V_{\mathrm ps}(\left|{\mathbf r -
   R}_i\right|).\eqnum{1}\label{1st}
\end{equation}
In the present work, we have developed a more physical pseudopotential than 
that used in earlier applications of CAPS \cite{caps1}; it will be discussed 
below.

The second approximation is that while the ions are treated
three-dimensionally, the electron density is
 restricted to cylindrical symmetry, i.e.,
in the solution of the electronic problem $V_{\mathrm ps}(\left|{\mathbf r -
   R}_i\right|)$ is replaced by its cylindrical average
\begin{equation}
\bar{V}_{\mathrm ps}(z,\rho;z_i,\rho_i)=
\frac{1}{2\pi}\int_0^{2 \pi}V_{\mathrm ps}(\left|
{\mathbf r-R}_i\right|)\,d\varphi. \eqnum{2}\label{2nd}
\end{equation}
This certainly is a simplification whose detailed consequences are
hard to judge {\it a priori}. However, since the photoabsorption cross
sections
of singly charged sodium clusters show that the electronic density of
most clusters has
an overall prolate, oblate or spherical shape \cite{mei,montag1,ep},
 this approximation
does not seem unreasonable for such systems.
A severe test for the quality of this
approximation will be a comparison of its results to those of
fully three dimensional methods.

Before we present the results of our calculations, we discuss the local
pseudopotential in some more detail. The most rigorous pseudopotentials in the
sense of Philipps and Kleinman \cite{philklein} and modern {\it ab initio}
pseudopotentials \cite{bachelet} are always non-local in the sense that 
each angular momentum component of
the valence electron feels a different potential. However, the use
of non-local pseudopotentials in the search for cluster
configurations quickly exhausts computational resources because of the
multiple projections that have to be done at every step of the calculation,
separately for each ion in the absence of any symmetry. Also, it has
been shown that some {\it ab initio} pseudopotentials do not necessarily
lead to a good agreement with experiment \cite{moullet1}. It therefore
makes sense to address the pseudopotential question from a more pragmatic 
point of view.

Already early in the development of pseudopotential theory, it has been
noted that by relaxing the Phillips-Kleinman condition, one can open
up a new class of pseudopotentials \cite{heineaba}.
They are called phenomenological pseudopotentials or model potentials
since they are
constructed by choosing some analytical function as a model potential and
adjusting its parameters to experimentally known quantities, e.g.,
an atomic energy level or some bulk properties. Such model potentials can
be nonlocal, or several partial-wave components may be chosen to be
the same \cite{heineaba}. For metals with a simple electronic structure
like that of sodium, one can in this way
construct pseudopotentials that are effectively local.
Various local pseudopotentials have, in fact, been successfully used
(see, e.g., Refs.\ \cite{ppstoer,landman,evc}).
But the question of how a valid local pseudopotential should be 
constructed is nontrivial. In Ref.\ \cite{evc} this question was addressed in
detail with an emphasis on solid-state properties. There, an ``evanescent
core potential'' was proposed, and we have used it in some test cases.
Most of our calculations, however, were done with a pseudopotential that
we have constructed especially for the use in finite sodium clusters,
as explained below.

Whereas in
solid-state physics an important criterion for the practical usefulness
of a pseudopotential is its fast convergence in reciprocal space, our aim is
its efficient handling in real space. We take up experience from CAPS 
\cite{caps1} and parameterize the pseudopotential on the basis of 
pseudodensities, related to (\ref{2nd}) via Poisson's equation,
which can be angle averaged analytically. This allows for 
a more
efficient solution of the Coulomb problem. Moreover, the short range of the
pseudodensities ensures a fast repositioning of the ions. Our pseudopotential
has the parameterization
\begin{displaymath}
V_{\mathrm ps}(r)=e^2\left\{\begin{array}{cl}
\frac{2 \pi}{3}\varrho_1 r^2 +c_1 & r<r_1 \nonumber \\
-\frac{q_1}{r}+\frac{2 \pi}{3}\varrho_2 r^2+c_2 & r_1 \leq r<
  r_2 \nonumber\\
-\frac{Z}{r} & r\geq r_2.
\end{array}
\right. 
\end{displaymath}
This corresponds to a pseudodensity with a two-step profile.
Four of the seven  parameters are fixed by requiring continuity of
$V_{ps}(r)$ and its derivative.
The remaining parameters $\varrho_1$,
$r_1$ and $r_2$ determine the physical properties of the pseudopotential.
Clusters contain from a few up to several thousand atoms, thus spanning
the region from the atom to the bulk material. Our aim is therefore to develop
a local pseudopotential that interpolates between atomic and bulk properties.
Thus we choose the parameters such that two quantities are reproduced
correctly: the atomic $3s$ energy level $e_{a}$ on one hand, and the 
bulk Wigner-Seitz radius $r_s$ on the other hand. The latter
is determined by the minimum of the
bulk energy per electron $e_{\mathrm b}$
in second order perturbation theory \cite{evc}.
Using $r_s=3.93 a_0$ and the
experimental value $e_a=-0.38$ Ry \cite{heineaba} fixes
$\varrho_1=-0.503 {a_0}^{-3}$
and $r_2=3.292 a_0$ and gives a constraint on $r_1$. The remaining freedom in
choosing $r_1$ was exploited to fit the bulk compressibility $B$ as closely
as possible to its
experimental value $B_{\mathrm exp}=0.073$ Mbar \cite{compress}, yielding
$B=0.0739$ Mbar for $r_1=0.641 a_0$.
With these parameters we obtain $e_{\mathrm b}=-6.20$ eV,
close to the experimental value
$-6.25$ eV. The interstitial density, defined as the difference between
the number of valence electrons in the Wigner-Seitz cell and in
the muffin-tin sphere \cite{evc},
takes the value $0.35$ with our pseudopotential.
This agrees within 3\% with the value given
in Ref.\ \cite{evc}. The band-structure energy is $0.15$ eV, in agreement
with Ref.\ \cite{evc}.

As a test-case study, we have calculated
the ground state of Na-7+ with our pseudopotential, with
the evanescent-core potential \cite{evc}, and with
the empty-core like pseudopotential \cite{caps1}.
The $D_{5h}$-geometry of Na7+ is well known from {\it ab initio} 
calculations \cite{bonacic4,martins1},
and we find the pentagonal bipyramid, shown in Fig.\ 1,
with all three pseudopotentials.
This demonstrates that CAPS can give realistic results even for
very small systems, and that the cylindrical averaging is not too
restrictive. The influence of the pseudopotential can be seen
in the bonding lengths, e.g.\ the distance between the two edges of the
bipyramid: our pseudopotential and the evanescent core
potential result in a distance of $6.03 a_0$ and $6.02 a_0$, respectively,
whereas the empty-core like pseudopotential leads to a shorter distance
of $5.64 a_0$. All these values lie in the range found in
fully three-dimensional calculations: Ref.\ \cite{martins1} quotes a distance
of $5.5 a_0$, Ref.\ \cite{bonacic3} one of $6.26 a_0$. 
\begin{figure}
\PostScript{0}{3.3}{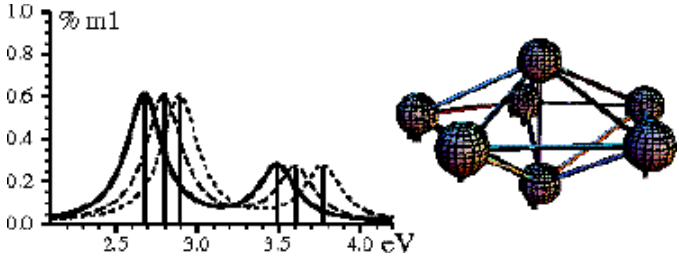}
\caption{Right-hand side: CAPS-geometry  of Na-7+. 
Left-hand side: percentages of the dipole $m_1$ sum rule obtained 
in the ``localized'' random-phase approximation
for different pseudopotentials. Full line: present
pseudopotential; dashed line: evanescent-core potential \cite{evc};
dotted line: empty-core like pseudopotential \cite{caps1}.
A phenomenological Lorentzian line broadening is applied.}
\label{na7p}
\end{figure}
Although these 
differences might appear small, they have noticeable consequences
for the photoabsorption spectrum, as shown in Fig.\ 1.
Here we plot the percentages of the energy-weighted dipole sum rule ($m_1$)
obtained in the ``localized'' random-phase approximation (LRPA) 
\cite{brack1,lrpa}. For all pseudopotentials, two dominant peaks 
are seen that together carry about 90 \% of the total
oscillator strength. Their relative heights
reflect the oblate deformation of the electron cloud in the 
$D_{5h}$ configuration.
(The remaining 10 \% of the dipole strength are scattered in
a region around 4.5 eV.)
The experimental photoabsorption spectrum \cite{expna7p} of Na7+ 
 shows these two peaks at 2.35
eV and 3.25 eV when measured at a temperature of 650 K; at 35 K the lower
transition splits into two close-lying peaks centered at 2.5 eV, while the
higher transition is shifted to 3.3 eV.
Our results do not reproduce the fine splitting of the lower peak. It is known
\cite{lrpa} that the LRPA cannot resolve splittings that are due to 
interference of the plasmon resonance with specific particle-hole
excitations, but it correctly reproduces the average peaks in the global
strength distribution. (The convergence of the LRPA basis is within 3\% for 
the peak positions.) 
The most important observation is, however, that all the
spectra are blueshifted with respect to the experiment, and that the
extent of the blueshift depends remarkably
on the pseudopotential employed in the calculation.
It ranges from 0.2 eV, found for the present pseudopotential,
to nearly 0.5 eV for the empty-core type, with the evanescent-core results
lying in between. That the empty-core like pseudopotential is the least
accurate is not astonishing, since its parameters were fitted
\cite{caps1} to the bulk value of $r_s$ only
in first-order perturbation theory. The discrepancy between the
evanescent-core pseudopotential and ours could be the consequence of
a better transferability of our pseudopotential; it might also
reflect the fact that the evanescent-core pseudopotential
is more difficult to handle numerically in our code.

A blueshift of the dipole resonance in sodium clusters with respect
to its experimental position
has been found in many calculations, and there
has been a long-standing discussion about its origin
\cite{guet1,revmod,ppstoer,bertsch,moullet2,isspic}. Our results show that
the detailed form of the
pseudopotential does strongly influence the resonance position, but that
other effects must also contribute to the discrepancy with experiment. 
We think that one important effect is the finite temperature, present in
most experiments, which results in an increased cluster volume, and thus to
a decrease of the plasma frequency of a few percent.
The experiments of Ref.\ \cite{expna7p} show, indeed, that
the average resonance positions are shifted to lower energies when the
temperature is increased.
\begin{figure}
\PostScript{0}{6}{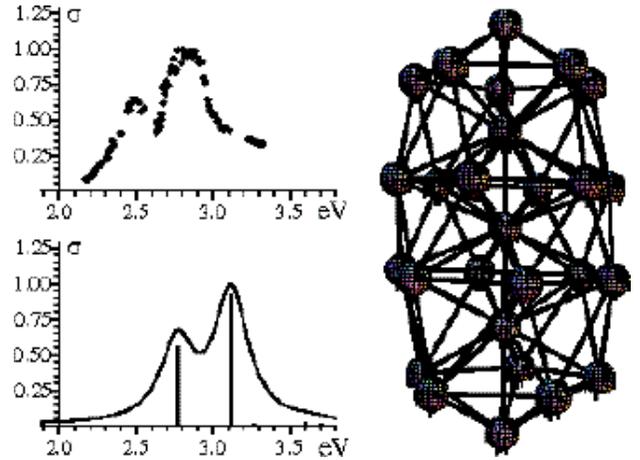}
\caption{Ionic structure and photoabsorption cross section $\sigma$ of Na-27+
in arbitrary units against eV. Upper panel left:
Experimental data \cite{private}. Lower panel left: LRPA photoabsorption 
spectrum. Right-hand side: CAPS structure.}
\label{pana27p22}
\end{figure}
This must be borne in mind when comparing the results of calculations
performed at T=0 to those experimental data of larger clusters, where
the photoabsorption was measured only at finite temperature.
On the other hand, the LDA leads to well known errors \cite{dreizler} that
also affect the optic response \cite{sic}.

\begin{figure}
\PostScript{0}{6}{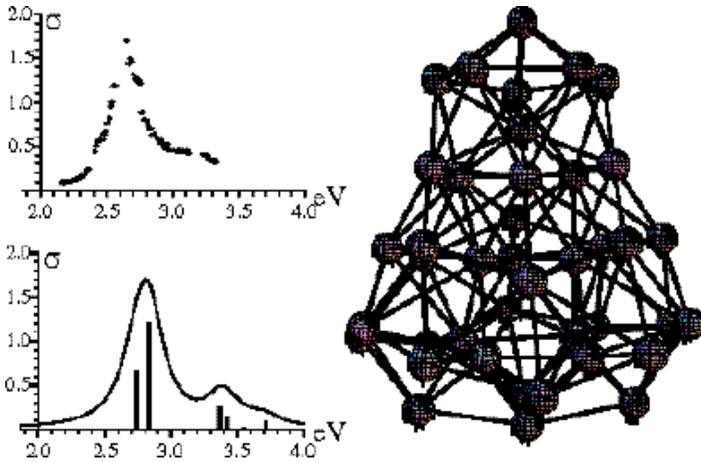}
\caption{Same as Fig.\ 1, but for Na-41+. Experimental data are from Ref.\ 
\cite{na41p}.} 
\label{pana41p}
\end{figure}

Encouraged by the correct prediction of the structure
of Na-7+, we have employed our scheme to calculate ground-state
structures and photoabsorption spectra of sodium clusters in a size
region where no self-consistent calculations with ionic structure
have been made so far. In Figures \ref{pana27p22} and \ref{pana41p}, 
we present the ionic geometries and photoabsorption spectra of Na-27+ 
and Na-41+. In both cases, we find good agreement with the
experimental results. The overall blueshift of the resonance peaks
of 7 - 9 \% can again be accounted for by the effects
mentioned above. The two pronounced peaks observed in Na-27+
are a consequence of an overall prolate arrangement of the ions. 
Together they exhaust $\sim$ 80\% of the total oscillator strength. 
The remaining strength is distributed around 4.3 eV, a region which 
was not scanned experimentally and thus omitted from the plot.

In contrast, Na-41+ has a ``magic'' configuration of 40 valence electrons,
leading to a nearly spherical density and almost no splitting of the
dipole resonance. The two strongest close-lying transitions
exhaust 64\% of the total oscillator strength. It is very interesting
to note that, besides 20\% of the strength lying outside the region of the 
experiment, another 16\% of the strength are scattered around 
3.4 eV. In the LRPA, these subpeaks are interpreted as internal compressional 
modes and modes of higher multipolarity that are coupled to the dipole 
oscillation by the ionic structure. Their contributions are seen to be
in good qualitative agreement with the high-energy shoulder observed 
in the experiment.

In summary, our studies have shown that CAPS, combined with the localized 
random-phase approximation, provide a powerful tool for calculating ionic 
geometries and photoabsorption spectra of medium-sized sodium
clusters where {\it ab initio} procedures are computationally not
possible with a full search of ionic structures. With a local pseudopotential
fitted to bulk and atomic properties, we reproduce the global features
of the dipole resonances, apart from a slight blue shift that leaves room for
temperature and non-LDA effects.
We believe that CAPS gains additional importance since its results
can serve as an input to three-dimensional calculations, thus
bringing closer an explanation of the recent observation of the melting
of clusters which depends
on a detailed knowledge of their ionic structure \cite{schmelzen}.

This work has been partially supported by the Deutsche Forschungsgemeinschaft.


\begin{references}

\bibitem{na41p}Th. Reiners {\it et al.},
Chem. Phys.\ Lett.\ {\bf 215}, 357 (1993).

\bibitem{expna7p}C. Ellert {\it et al.}, Phys.\ Rev.\ Lett.\ 75, 1731 (1995).

\bibitem{mei}P. Meibom {\it et al.}, Z.\ Phys.\ D {\bf 40}, 258 (1997);
             J. Borggreen {\it et al.}, Phys.\ Rev.\ B {\bf 48}, 17507 (1995). 

\bibitem{ekardt1}W. Ekardt, Phys.\ Rev.\ Lett.\ {\bf 52}, 1925 (1984).

\bibitem{beck}D. E. Beck, Solid State Commun. {\bf 49}, 381 (1984).

\bibitem{brack1}M. Brack, Phys. Rev.\ B {\bf 39}, 3533 (1989).

\bibitem{yan1}C. Yannouleas {\it et al.},
 Phys.\ Rev.\ Lett.\ {\bf 63}, 255 (1989).

\bibitem{montag1}Th. Hirschmann, M. Brack, and P.-G. Reinhard, 
 Z.\ Phys.\ D {\bf 40}, 254 (1997).

\bibitem{guet1}M. Madjet, C. Guet, and W. Johnson, Phys.\ Rev.\ A {\bf 51}, 
1327 (1995).

\bibitem{revmod}M. Brack, Rev.\ Mod.\ Phys.\ {\bf 65}, 677 (1993).

\bibitem{bonacic3}V. Bona\v{c}ic-Kouteck\'{y}, P. Fantucci and J.
  Kouteck\'{y}, Chem.\ Rev.\ {\bf 91}, 1035 (1991).

\bibitem{bonacic4}V. Bona\v{c}ic-Kouteck\'{y} {\it et al.}, J.\ Chem.\ Phys.\
   {\bf 104}, 1427 (1996).

\bibitem{carpar}R. Car and M. Parrinello, Phys.\ Rev.\ Lett.\ {\bf 55},
  2471 (1985).

\bibitem{rothlis}U. R\"othlisberger and W. Andreoni, J.\ Chem.\ Phys.\
  {\bf 94}, 8129 (1991).

\bibitem{martins1}J. L. Martins, J. Buttet, and R. Car, Phys.\ Rev.\ B 
{\bf 31}, 1804 (1985).

\bibitem{iniguez}M. P. I\~{n}iguez {\it et al.},
  Z.\ Phys.\ D {\bf 11}, 163 (1989).

\bibitem{borstel}G. Borstel {\it et al.},
  in {\it Lecture Notes in Physics}, {\em Nuclear Physics Concepts in the Study of
Atomic Cluster Physics}, edited by R. Schmidt, H. O. Lutz
  and R. Dreizler (Springer, Berlin, 1992), Vol. 404.

\bibitem{serra}Ll. Serra {\it et al.}, Phys.\ Rev.\ B {\bf 48}, 14708 (1993).

\bibitem{alasia}F. Alasia {\it et al.}, Phys.\ Rev.\ B {\bf 52}, 8488 (1995).

\bibitem{ppstoer}W.-D. Sch\"one, W. Ekardt and J. M. Pacheco, Z.\ Phys.\ D 
 {\bf 36}, 65 (1996).

\bibitem{bertsch}
K. Yabana and G.F. Bertsch, Phys.\ Rev.\ B {\bf 54}, 4484 (1996).

\bibitem{spiegelmann}R. Poteau and F. Spiegelmann, J.\ Chem.\ Phys.\ {\bf 98},
                     6540 (1993).

\bibitem{caps1}B. Montag and P.-G. Reinhard, Z.\ Phys.\ D {\bf 33}, 265 (1995).

\bibitem{dreizler}R. M. Dreizler and E. K. U. Gross, {\em Density Functional
  Theory } (Springer, Berlin 1990).

\bibitem{pw}J. P. Perdew and Y. Wang, Phys.\ Rev.\ B {\bf 45}, 13244 (1992).

\bibitem{ep}W. Ekardt and Z. Penzar, Phys.\ Rev.\ B {\bf 43}, 1322, (1991).

\bibitem{philklein}J. C. Phillips and L. Kleinman, Phys.\ Rev.\ {\bf 116}, 287
                   (1959).

\bibitem{bachelet}G. B. Bachelet, D. R. Haman, and M. Schl\"uter, 
Phys.\ Rev.\ B {\bf 26}, 4199 (1982).

\bibitem{moullet1}I. Moullet and J. L. Martins, J.\ Chem.\ Phys.\ {\bf 92},
  527 (1990).

\bibitem{heineaba}I. V. Abarenkov and V. Heine, Phil.\ Mag.\ {\bf 12}, 529
 (1965).

\bibitem{landman}R. N. Barnett, U. Landman, and C. L. Cleveland, Phys.\ Rev.\ B
{\bf 27}, 6534 (1983)

\bibitem{evc}C. Fiolhais {\it et al.},
Phys. Rev. B {\bf 51}, 14001 (1995); and Phys.\ Rev.\ B
 {\bf 53}, 13193 (1996).
\bibitem{compress}M. S. Anderson and C. A. Swenson, Phys.\ Rev.\ B {\bf 28},
  5395 (1983).

\bibitem{moullet2}I. Moullet {\it et al.},
  Phys.\ Rev.\ B {\bf 42}, 11589 (1990).

\bibitem{isspic}P.-G. Reinhard {\it et al.}, Z.\ Phys.\ D {\bf 40}, 314 (1997).

\bibitem{lrpa}P.-G. Reinhard, M. Brack and O. Genzken, Phys. Rev. A {\bf 41},
  5568 (1990); P.-G. Reinhard, O. Genzken, and M. Brack, Ann.\ Phys.\ 
  (Leipzig) {\bf 51}, 576 (1996).

\bibitem{private}M. Schmidt and H. Haberland, private communication.

\bibitem{sic} See sect. IV.B.1 of Ref.\ \cite{revmod} for a critical
 discussion of the so called self-interaction correction and other extensions
of the LDA in this context.

\bibitem{schmelzen}M. Schmidt {\it et al.},
 Phys.\ Rev.\ Lett.\ {\bf 79}, 99 (1997).

\end{references}
\end{document}